\documentclass[aip,reprint,amsmath,amssymb]{revtex4-1}

\usepackage{graphicx} 
\usepackage{dcolumn}
\usepackage{bm}
\usepackage{amsmath}

\draft % marks overfull lines with a black rule on the right

\begin{document}

\title{Functions to map photoelectron distributions in a variety of setups in angle-resolved photoemission spectroscopy}

\author{Y.~Ishida}
\email[]{ishiday@issp.u-tokyo.ac.jp}
\author{S.~Shin}
\affiliation{ISSP, University of Tokyo, 5-1-5 Kashiwa-no-ha, Kashiwa, Chiba 277-8581, Japan}

\date{\today}

\begin{abstract}
The distribution of photoelectrons acquired in angle-resolved photoemission spectroscopy can be mapped onto energy-momentum space of the Bloch electrons in the crystal. The explicit forms of the mapping function $f$ depend on the configuration of the apparatus as well as on the type of the photoelectron analyzer. We show that the existence of the analytic forms of $f^{\text{-}1}$ is guaranteed in a variety of setups. The variety includes the case when the analyzer is equipped with a photoelectron deflector. Thereby, we provide a demonstrative mapping program implemented by an algorithm that utilizes both $f$ and $f^{\text{-}1}$. The mapping methodology is also usable in other spectroscopic methods such as momentum-resolved electron-energy loss spectroscopy. 
\end{abstract}

\pacs{}

\maketitle %\maketitle must follow title, authors, abstract and \pacs
\draft % marks overfull lines with a black rule on the right

\section{Introduction}
\label{Intro}

Band structures of crystals can be visualized by using angle-resolved photoemission spectroscopy (ARPES)~\cite{80PRB_Chiang}. The visualization procedure is based on the principle that the kinetic energy ($\varepsilon_{\rm kin}$) and angular distribution of photoelectrons can be mapped onto energy ($\omega$) and momentum space of Bloch electrons in the crystal~\cite{83AdvPhys_Himpsel}. The well-established methodology makes ARPES a powerful tool to study the electronic structures of crystals~\cite{03RMP_ARPES}. 

\begin{figure} [htb]
\includegraphics[width=\columnwidth]{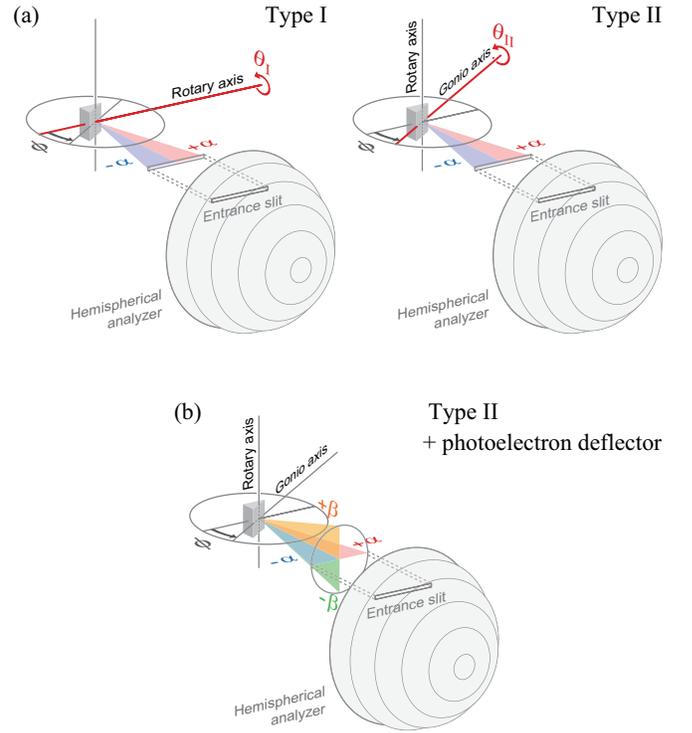}
\caption{\label{fig1} ARPES setups. (a) Two types of roto-axes configurations. Type I (left): The manipulator rotary axis is parallel to the entrance slit of a hemispherical analyzer. Type II (right): The manipulator rotary axis is perpendicular to the slit. A two dimensional angular distribution is obtained by rotating the sample around the axis colored in red; as a result, the distribution is recorded in the $\alpha$\,-\,$\theta_{\rm I}$($\alpha$\,-\,$\theta_{\rm II}$) plane for type I (type II). (b) Type II equipped with a photoelectron deflector. The two-dimensional distribution is recorded in the $\alpha$\,-\,$\beta$ plane.}
\end{figure}

The explicit forms of the mapping function, or the way the angular variables appear in the function, depend on the configuration of the ARPES apparatus. In order to illustrate the dependency, we show, in Fig.~\ref{fig1}(a), two typical roto-axes configurations with respect to the hemispherical analyzer that has a slit-type aperture. In type I (type II), the rotary axis of the manipulator that holds the sample is parallel (perpendicular) to the direction of the slit; and when acquiring a two-dimensional angular distribution of the photoelectrons, the sample is rotated step by step around that rotary axis (another axis often called ``gonio"~\cite{03RSI_Aiura}). Because the axis of rotation during the acquisition is inequivalent between the two, the forms vary with the configuration. The forms change further when the hemispherical analyzer is updated to state-of-the-art equipped with a photoelectron deflector; see, Fig.~\ref{fig1}(b). The analyzer equipped with a deflector can also detect photoelectrons directed off the slit, and achieves the so-called slit-less concept. In such a setup, a new angular variable $\beta$ has to be taken into account explicitly, because $\beta$ is independent of the angles that describe the sample orientation. 

Thus, in order to map the ARPES data onto momentum space, the explicit forms of the mapping function $f$ is needed for the particular setup of the apparatus. Practically, knowing the forms of the inverse mapping function is also very useful. A mapping algorithm can be made that utilizes both $f$ and $f^{\text{-}1}$, and such an algorithm can shorten the computation time for the mapping compared to the case when only $f$ is used; see, Appendix~\ref{appendix}. However, the derivation of the explicit forms can be complicated particularly when the number of the angles that should be considered in the setup becomes large. 

In the present article, we systematically investigate the derivation of the explicit forms of $f^{\text{-}1}$ for a variety of setups. The variety includes the case when the analyzer is equipped with a deflector. We explicate the underlying mathematical reasons for the existence of the analytic solutions, and guarantee their existence in the variety. That is, we warrant that the mapping program implemented by both $f$ and $f^{\text{-}1}$ can be written for a number of setups. For practical usage, we provide the explicit forms for some typical setups including those illustrated in Fig.~\ref{fig1}, and also demonstrate a mapping program. 

While the focus of the present article is on ARPES, the mapping methodology described herein is also applicable to other spectroscopic methods such as momentum-resolved electron-energy loss spectroscopy, the technique of which is also developing rapidly~\cite{15RSI_EELS,17Science_EELS}. 

The article is structured as follows. In section \ref{sec_typeI}, we show an instructional derivation of the analytic forms of $f$ and $f^{\text{-}1}$ for type I. Then in sections \ref{sec_typeII} and \ref{sec_deflector}, we consider respectively the case for type II and the case when the deflector is equipped. In section \ref{sec_systematics}, we extract the systematics in the derivations and investigate the reason why the analytic solutions can exist. Discussion is provided in section \ref{sec_discussion}. In Appendix, we summarize the analytic forms of $f$ and $f^{\text{-}1}$ for some typical setups (Appendix~\ref{appendix}), and also provide some tips for the angular notation when the deflector-type analyzer is used (Appendix~\ref{appendix_notation}). The mapping program is provided in Supplementary Material. 

When there is no confusion, we abbreviate sine and cosine functions as follows: $\cos \beta\,\to\,{\rm c}\beta$; $\sin \beta\,\to\,{\rm s}\beta$. 

\begin{figure*} [htb]
\includegraphics[width=\textwidth]{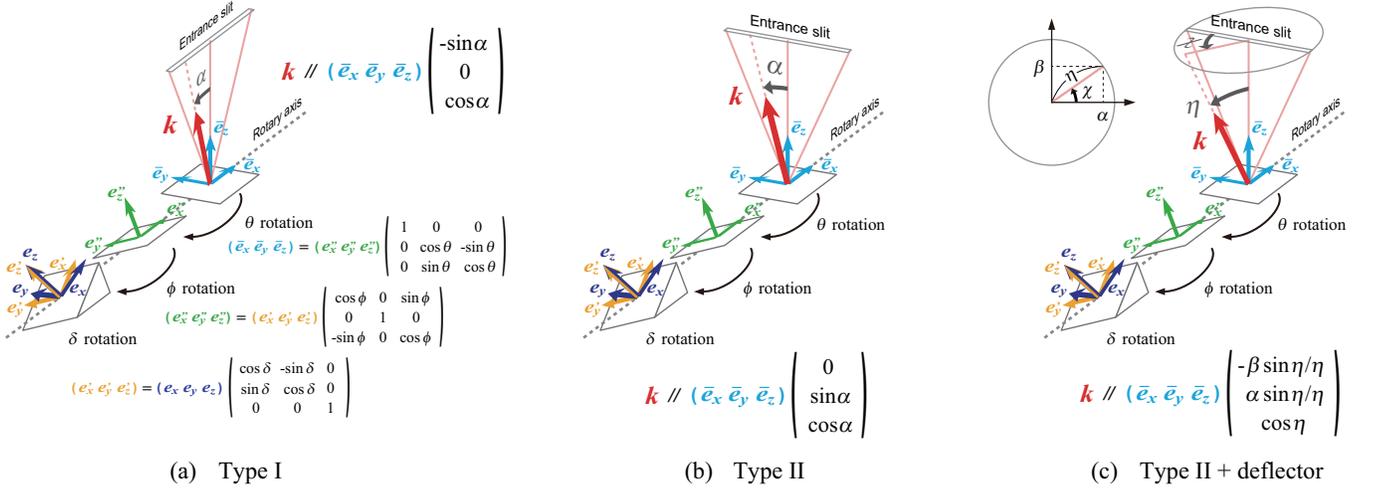}
\caption{\label{fig2} Definition of the angles and rotations for type I (a), type II (b), and type II equipped with a deflector (c). The basis colored in sky blue is fixed on the analyzer's frame; with that basis, the photoelectron momentum \mbox{\boldmath $k$} is easily describable. The basis colored in dark blue is fixed on the sample surface; with that basis, we need to describe \mbox{\boldmath $k$}. The sky-blue basis is reached from the dark-blue basis by successively applying rotation operations of angles $\delta$, $\phi$, and $\theta$ around the $z$, $y$, and $x$ axes, respectively. In (c), we also illustrated the relationship between the polar/azimuth angles and the polar-angular notation. }
\end{figure*}

\section{The case for type I}
\label{sec_typeI}

Figure~\ref{fig2}(a) illustrates the type I configuration. Four Cartesian bases are introduced. Among the four, $(\mbox{\boldmath $e$}_x\,\mbox{\boldmath $e$}_y\,\mbox{\boldmath $e$}_z)$ has its $x$\,-\,$y$ plane fixed on the crystal surface, and $(\mbox{\boldmath $\bar{e}$}_x\,\mbox{\boldmath $\bar{e}$}_y\,\mbox{\boldmath $\bar{e}$}_z)$ is fixed to the analyzer's frame. $\alpha$ is the emission angle of the photoelectron with respect to the slit, $\theta$ is the angle that is varied step by step during the data acquisition, and $\phi$ and $\delta$ are angular parameters. 

First, we derive the mapping function from the space for the photoelectron to the space for the Bloch electron; namely, to describe ($\omega$, $k_x$, $k_y$) with the measurable ($\varepsilon_{\rm kin}$, $\alpha$, $\theta$). Here, $k_x$ and $k_y$ are the momentum components of the Bloch electron parallel to the crystal surface. 

The energy sector of the mapping function is simple: 
\begin{equation}
\label{eq_energy}
\omega(\varepsilon_{\rm kin}) = \varepsilon_{\rm kin} - h\nu + \phi_{\rm w}, 
\end{equation}
where $h\nu$ and $\phi_{\rm w}$ are parameters that correspond to the photon energy and work function, respectively. 

The problem for the momentum sector is equivalent to knowing the components of the photoelectron momentum \mbox{\boldmath $k$} projected on the crystal surface, because those are preserved to ($k_x, k_y$) of the Bloch electron. We thus need to write down \mbox{\boldmath $k$} with (\mbox{\boldmath $e$}$_x$\,\mbox{\boldmath $e$}$_y$\,\mbox{\boldmath $e$}$_z$); somehow, \mbox{\boldmath $k$} is conveniently described with (\mbox{\boldmath $\bar{e}$}$_x$\,\mbox{\boldmath $\bar{e}$}$_y$\,\mbox{\boldmath $\bar{e}$}$_z$): That is,  
\begin{eqnarray}
	\mbox{\boldmath $k$} = (\mbox{\boldmath $e$}_{x}\,\mbox{\boldmath $e$}_{y}\,\mbox{\boldmath $e$}_{z}) \left(
	\begin{array}{r}
		k_x \\
		k_y \\
		k_z
	\end{array}
	\right) =  k (\mbox{\boldmath $\bar{e}$}_{x}\,\mbox{\boldmath $\bar{e}$}_{y}\,\mbox{\boldmath $\bar{e}$}_{z}) \left(
    \begin{array}{r}
		-\rm{s}\alpha \\
		0 \\
		\rm{c}\alpha
    \end{array}
	\right), 
\end{eqnarray} 
where $k(\varepsilon_{\rm kin}) = \sqrt{2mc^2\varepsilon_{\rm kin}}/\hbar c$ ($mc^2$ = 511~keV and $\hbar c$ = 1970~eV\AA\, are the constants). Therefore, we need to know the relationship between the two bases, which is 
\begin{equation}
(\mbox{\boldmath $\bar{e}$}_{x}\,\mbox{\boldmath $\bar{e}$}_{y}\,\mbox{\boldmath $\bar{e}$}_{z})  = 
(\mbox{\boldmath $e$}_x\,\mbox{\boldmath $e$}_y\,\mbox{\boldmath $e$}_z) T_{\rm rot},
\end{equation}
where
\begin{eqnarray}
T_{\rm rot} = \left(
\begin{array}{ccc}
\rm{c}\phi \rm{c}\delta	& -\rm{c}\theta \rm{s}\delta + \rm{s}\theta \rm{s}\phi \rm{c}\delta 	&	\rm{s}\theta \rm{s}\delta +  \rm{c}\theta \rm{s}\phi \rm{c}\delta \\
\rm{c}\phi \rm{s}\delta	&	\rm{c}\theta \rm{c}\delta + \rm{s}\theta \rm{s}\phi \rm{s}\delta 	&	-\rm{s}\theta \rm{c}\delta + \rm{c}\theta \rm{s}\phi \rm{s}\delta \\
-\rm{s}\phi						&	\rm{s}\theta \rm{c}\phi 																	&	\rm{c}\theta \rm{c}\phi 
\end{array}  
\right).
\end{eqnarray}
Altogether, 
\begin{eqnarray}
\label{eq_mapping}
\left(
\begin{array}{c}
k_x \\
k_y \\
k_z
\end{array}
\right) = k(\varepsilon_{\rm kin})T_{\rm rot}(\theta)
\left(
\begin{array}{r}
-\rm{s}\alpha \\
0	\\
\rm{c}\alpha
\end{array}  
\right). 
\end{eqnarray}
The first and second lines of the equation are the forms of the mapping function for the momentum sector. 

Next, we derive the inverse mapping function; namely to describe ($\varepsilon_{\rm kin}$, $\alpha$, $\theta$) with the variables ($\omega$, $k_x$, $k_y$). 

The inverse function for the energy sector is 
\begin{equation}
\varepsilon_{\rm kin}(\omega) = \omega + h\nu - \phi_{\rm w}.  \label{eq_energy_inv}
\end{equation}

As for the angular sector, the first step is to describe the photoelectron's momentum by using the variables for the Bloch electrons. This can be done on the sample's basis as follows: 
\begin{eqnarray}
\mbox{\boldmath $k$} = (\mbox{\boldmath $e$}_x\,\mbox{\boldmath $e$}_y\,\mbox{\boldmath $e$}_z) \left(
\begin{array}{c}
k_x \\
k_y \\
%\{k(\omega)^2 - k_x^2 - k_y^2\}^{1/2} 
k_z(\omega, k_x, k_y)
\end{array}
\right).
\end{eqnarray}
Here, $k_z (\omega, k_x, k_y) = \sqrt{k^2(\omega) - k_x^2 - k_y^2}$, and $k(\omega) = \sqrt{2mc^2 (\omega + h\nu - \phi_{\rm w})}/\hbar c$. Then, we rotate the sample step by step with respect to the analyzer, and search the angle $\theta$ when the photoelectron enters the silt. Note that $\delta$ and $\phi$ are fixed during the rotation. The momentum components seen from the analyzer frame are 
\begin{eqnarray}
\label{eq_reorient}
\left(
	\begin{array}{c}
	{\bar k_x}(\theta) \\
	{\bar k_y}(\theta) \\
	{\bar k_z} (\theta)
	\end{array}
\right) = T_{\rm rot}^{\text{-}1}(\theta) \left(
\begin{array}{c}
k_x \\
k_y \\
%\{k(\omega)^2 - k_x^2 - k_y^2\}^{1/2} 
k_z(\omega, k_x, k_y)
\end{array}
\right),
\end{eqnarray}
where 
\begin{eqnarray}
T_{\rm rot}^{\text{-}1} = \left(
\begin{array}{ccc}
\rm{c}\phi \rm{c}\delta	& \rm{c}\phi \rm{s}\delta	&	-\rm{s}\phi \\
\rm{s}\theta \rm{s}\phi \rm{c}\delta - \rm{c}\theta \rm{s}\delta		&	\rm{s}\theta \rm{s}\phi \rm{s}\delta + \rm{c}\theta \rm{c}\delta	&	\rm{s}\theta \rm{c}\phi \\
\rm{c}\theta \rm{s}\phi \rm{c}\delta + \rm{s}\theta \rm{s}\delta		&	\rm{c}\theta \rm{s}\phi \rm{s}\delta - \rm{s}\theta \rm{c}\delta	&	\rm{c}\theta \rm{c}\phi 
\end{array}  
\right).
\end{eqnarray}
The condition for the photoelectron to enter the slit is
\begin{equation}
{\bar k_y}(\theta) = 0. \label{eq_proper}
\end{equation}
The left hand side of the entrance condition has the form $A \cos\theta - B \sin\theta$, or is a linear form of $\cos\theta$ and $\sin\theta$, where $A$ and $B$ are independent  parameters of $\theta$. Therefore, the solution exists in the form $\theta = \tan^{\text{-}1} (A/B)$. Explicitly,  
\begin{equation}
\label{eq_inv_theta}
\theta = \tan^{\text{-}1}\left(\frac{\rm{s}\delta \mbox{$k_x$}  - \rm{c}\delta \mbox{$k_y$}}{\rm{s}\phi \rm{c}\delta \mbox{$k_x$} + \rm{s}\phi \rm{s}\delta \mbox{$k_y$} + \rm{c}\phi \mbox{$k_z$}}\right).  
\end{equation}
%where $k_z (\omega, k_x, k_y) = \sqrt{k^2(\omega) - k_x^2 - k_y^2}$. 
When the photoelectron is directed toward the slit, the emission angle $\alpha$ can be solved by using the match of ${\bar k_x}$ to $-k\sin\alpha$:
\begin{equation}\label{eq_inv_matching_I}
{\bar k_x}(\theta) = -k\sin\alpha. 
\end{equation}
By operating $\sin^{\text{-}1}$ on the matching condition, we obtain 
\begin{equation}
\alpha = \sin^{\text{-}1}\left(\frac{-\rm{c}\phi \rm{c}\delta\mbox{$k_x$} - \rm{c}\phi \rm{s}\delta\mbox{$k_y$} + \rm{s}\phi\mbox{$k_z$}}{\mbox{$k$}}\right). \label{eq_inv_alpha} 
\end{equation}
Equations (\ref{eq_inv_theta}) and (\ref{eq_inv_alpha}) are the forms of the inverse mapping function for the angular sector. 

\section{The case for type II}
\label{sec_typeII}

In the case for type II, the slit is directed perpendicular to the rotary axis. Thus, the components of \mbox{\boldmath $k$} that is accepted by the slit are changed from the type I case. In addition, the angle varied during the data acquisition changes from $\theta$ to $\phi$. In other words, the role of being a variable or a parameter is exchanged between $\theta$ and $\phi$.  

With these in mind, the equation that corresponds to Eq.~(\ref{eq_mapping}) of type I becomes [see, Fig.~\ref{fig2}(b)]
\begin{eqnarray} 
\label{eq_mapping_II}
\left(
\begin{array}{c}
k_x \\
k_y \\
k_z
\end{array}
\right) = k(\varepsilon_{\rm kin})T_{\rm rot}(\phi)
\left(
\begin{array}{c}
0 \\
\rm{s}\alpha \\
\rm{c}\alpha
\end{array}  
\right), 
\end{eqnarray}
and the forms of the mapping functions for the momentum sector are read from its first and second lines. The form for the energy sector is the same to Eq.~(\ref{eq_energy}). 

As for the inverse mapping function, the equation that describes the rotation of the sample appears the same to Eq.~(\ref{eq_reorient}), but we remind that the rotation is done by varying $\phi$, while the parameters $\delta$ and $\theta$ are fixed. The entrance and matching conditions, which respectively correspond to Eqs.~(\ref{eq_proper}) and (\ref{eq_inv_matching_I}) of type I, are as follows:   
\begin{eqnarray}
{\bar k_x}(\phi) & = & 0;  \\ 
{\bar k_y}(\phi) & = & k \sin\alpha.  
\end{eqnarray}
By solving the conditions, we obtain the forms of the inverse mapping function for the angular sector: 
\begin{eqnarray}
\phi & = &\tan^{\text{-}1}\left(\frac{\rm{c}\delta \mbox{$k_x$}  + \rm{s}\delta \mbox{$k_y$}}{\mbox{$k_z$}}\right);  \label{eq_phi_inv_IIa}  \\
\alpha &= & \sin^{\text{-}1} \left[ 
\begin{array}{c}    
( \rm{s}\theta \rm{s}\phi \rm{c}\delta - \rm{c}\theta\rm{s}\delta ) \times \mbox{$k_x / k$} \\ 
+ (\rm{s}\theta \rm{s}\phi \rm{s}\delta + \rm{c}\theta \rm{c}\delta) \times \mbox{$k_y / k$} \\
+ (\rm{s}\theta \rm{c}\phi) \times \mbox{$k_z / k$} 
\end{array}
\right]. \label{eq_phi_inv_IIb}
\end{eqnarray}
Here, $k$ and $k_z$ as well as $\phi$ are functions of $(\omega, k_x, k_y)$. Equations (\ref{eq_phi_inv_IIa}) and (\ref{eq_phi_inv_IIb}) clearly demonstrate that the explicit forms cannot be obtained just by exchanging $\theta$ and $\phi$ in those of type I, Eqs.~(\ref{eq_inv_theta}) and (\ref{eq_inv_alpha}), or that rotation operations do not commute. The forms of Eq.~(\ref{eq_phi_inv_IIb}) can be simplified by using $\sin(\tan^{\text{-}1}x) = x/\sqrt{1 + x^2}$ and $\cos(\tan^{\text{-}1}x) = 1/\sqrt{1 + x^2}$; see, Appendix~\ref{appendix}. As for the energy sector, the form of the inverse function is not changed from Eq.~(\ref{eq_energy_inv}).

\section{The case with a deflector}
\label{sec_deflector} 

When the hemispherical analyzer that has a slit is further equipped with a potoelectron deflector, two dimensional angular distributions can be obtained without changing the orientation of the sample.  A pair of angular variables ($\alpha$, $\beta$) specifies the direction of the photoelectron momentum, while the set of parameters ($\theta$, $\phi$, $\delta$) fixes the crystal orientation. Our goal is to describe ($\omega, k_x, k_y$) by ($\varepsilon_{\rm kin}, \alpha, \beta$) and {\it vice versa}. The derivation shown below starts without explicating the direction of the slit, thanks to the slit-less concept achieved when the deflector is equipped; the explication will be done at the end of the section. 

The so-called polar angular notation is a convenient way to describe the direction of the photoelectron, and is adopted in state-of-the-art analyzers~\cite{Communication}. The notation is described in the upper left of Fig.~\ref{fig2}(c); also see, Appendix~\ref{appendix_notation}. The photoelectron momentum can be described by the two angular variables ($\alpha$, $\beta$) in the analyzer's frame as
\begin{eqnarray}
\mbox{\boldmath $k$} = k(\varepsilon_{\rm kin}) (\mbox{\boldmath $\bar{e}$}_{x}\,\mbox{\boldmath $\bar{e}$}_{y}\,\mbox{\boldmath $\bar{e}$}_{z})
\left(
\begin{array}{c}
-\beta\rm{s}\mbox{$\eta/\eta$} \\
\alpha\rm{s}\mbox{$\eta/\eta$} \\
 \rm{c}\mbox{$\eta$}
\end{array}
\right), 
\end{eqnarray}
where $\eta = \sqrt{\alpha^2 + \beta^2}$. Thus, the mapping function for the angular sector is read from the first and second lines of the following equation: 
\begin{eqnarray}
\label{eq_mapping_DA}
\left(
\begin{array}{c}
k_x \\
k_y	\\
k_z
\end{array}
\right) = k(\varepsilon_{\rm kin})T_{\rm rot}
\left(
\begin{array}{c}
-\beta\rm{s}\mbox{$\eta/\eta$} \\
\alpha\rm{s}\mbox{$\eta/\eta$} \\
 \rm{c}\mbox{$\eta$}
\end{array}
\right).
\end{eqnarray}
Note, the set of the polar angles ($\alpha$, $\beta$) is difficult to be illustrated in the real space, but can be in the parametric space, as shown in Fig.~\ref{fig2}(c); also see, Appendix~\ref{appendix_notation}.  

In order to derive the forms of $f^{\text{-}1}$, we first describe the photoelectron momentum by using the variables set for the Bloch electron ($\omega, k_x, k_y$) in the sample's frame, and then rewrite the components in the analyzer's frame, as done in Eq.~(\ref{eq_reorient}). Subsequent procedure becomes conceptually simpler than the former cases, because there is no need to rotate the sample any more. We need to know the angular variables for the momentum vector as is. That is, $T_{\rm rot}$ is a constant matrix, and we solve 
\begin{eqnarray}
\left(
	\begin{array}{c}
	{\bar k_x} \\
	{\bar k_y} \\
	{\bar k_z}
	\end{array}
\right) = T_{\rm rot}^{\text{-}1} \left(
\begin{array}{c}
k_x \\
k_y \\
k_z 
\end{array}
\right) = k(\omega)\left(
\begin{array}{c}
-\beta\rm{s}\mbox{$\eta/\eta$} \\
\alpha\rm{s}\mbox{$\eta/\eta$} \\
 \rm{c}\mbox{$\eta$}
\end{array}
\right)
\end{eqnarray}
for $\alpha$ and $\beta$. Their solutions exist as follows:  
\begin{eqnarray}
\alpha & = & \frac{{\bar k_y}}{\sqrt{k^2 - {\bar k_z}^2}}\cos^{\text{-}1}\left(\frac{{\bar k_z}}{k}\right);	\label{eq_final1} \\
\beta & = & \frac{-{\bar k_x}}{\sqrt{k^2 - {\bar k_z}^2}}\cos^{\text{-}1}\left(\frac{{\bar k_z}}{k}\right).	\label{eq_final2} 
\end{eqnarray}
Here we used $\sin(\cos^{\text{-}1}x) = \sqrt{1 - x^2}$. The inverse functions include $k$ and $k_z$ that are functions of ($\omega, k_x, k_y$) and contain the parameters ($h\nu$, $\phi_{\rm W}$, $\theta$, $\phi$, $\delta$). The existence of the solutions owes to the nature of the polar-angular notation; see the contrasted description after Eq.~(\ref{eq_proper}). 

If we regard that the photoelectrons are directed towards the slit when $\beta = 0$, Eqs.~(\ref{eq_mapping_DA}) - (\ref{eq_final2}) are the forms for the type II configuration. Those for type I are obtained by switching $(\alpha, \beta)$ to $(-\beta, \alpha)$ in the equations. Because the angular parametric space spanned by $(\alpha, \beta)$ can be rotated independent of $\theta$, $\phi$, and $\delta$, thanks to the polar-angular notation, the principal axis of the parametric space can be taken in any direction. In other words, when the analyzer is rotated around the electron-lens axis, the ARPES image seen in the angular space just rotates without deformation. Thus, the forms can also be used for setups that has the silt-less-concept analyzer such as the display-type analyzer~\cite{88RSI_Daimon} and time-of-flight-type analyzer equipped with a two-dimensional detector~\cite{11PRL_WangGedik}.

\section{Systematic treatment}
\label{sec_systematics}

Having considered the three cases in sections~\ref{sec_typeI}\,-\,\ref{sec_deflector}, we here extract the systematics when deriving the explicit forms of $f$ and $f^{\text{-}1}$, and investigate the reason why the analytic forms of $f^{\text{-}1}$ can exist. We show that the reason for the existence differs between the cases for hemispherical analyzers that have a slit and those that achieve the silt-less concept. The generalization would also be useful when developing a mapping program that can handle the datasets recorded under a variety of setups. 

\begin{figure*} [htb]
\includegraphics[width=\textwidth]{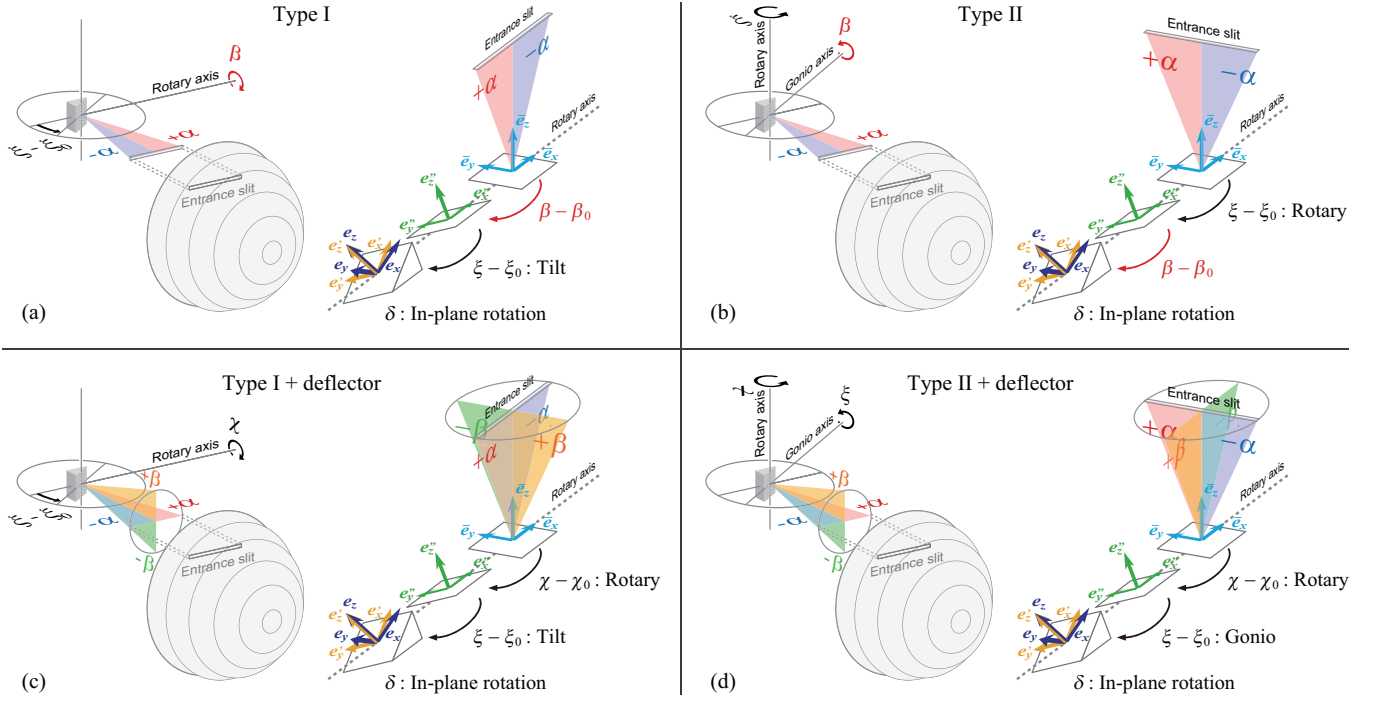}
\caption{\label{fig3} ARPES setups. (a) Type I: The rotary axis is parallel to the direction of the analyzer slit. (b) Type II: The rotary axis is perpendicular to the slit direction. (c) Type I + deflector: The axis configuration is identical to type I, and the analyzer is equipped with a photoelectron deflector. (d) Type II + deflector: The axis configuration is that of type II, and the analyzer has a deflector. The type I setting corresponds, for example, to those of laser ARPES~\cite{08RSI_Kiss} and time-resolved ARPES apparatuses~\cite{14RSI_Ishida} in ISSP, University of Tokyo; type II corresponds to those of SAMRAI in UVSOR~\cite{10RSI_UVSOR} and of ESPRESSO in HiSOR~\cite{11RSI_HiSOR}; and type II equipped with a deflector is found in the spin-resolved ARPES apparatus at ISSP, University of Tokyo~\cite{16RSI_Yaji}. 
}
\end{figure*}

The primary difference among the three cases was in the pair of the angular variables. Raw ARPES data were recorded in the parametric space of ($\alpha, \theta$), ($\alpha, \phi$), and ($\alpha, \beta$) for type I, type II, and type II with a deflector, respectively. In order to eliminate the apparent difference, we rename the angles so that all the raw ARPES data are spanned by ($\alpha, \beta$); see, Fig.~\ref{fig3} and Appendix~\ref{appendix}. Here, $\alpha$ refers to the emission angle along the direction of the slit, and $\beta$ refers to the angle perpendicular to $\alpha$ in the parametric space. Other angles are reassigned to $\delta$, $\xi$, $\ldots$ and are treated as parameters similar to $h\nu$ and $\phi_{\rm W}$.

After the renaming of the angles, the problem is reduced to describing ($\omega, k_x, k_y$) by the variables ($\varepsilon_{\rm kin}, \alpha, \beta$) and {\it vice versa}, while ($\delta$, $\xi$, $\ldots$, $h\nu$, $\phi_{\rm W}$) are treated as parameters.  

As for the mapping functions for the angular sector, the forms are combined into 
\begin{eqnarray}
\label{eq_mapping_sys}
\left(
\begin{array}{c}
k_x \\
k_y \\
k_z
\end{array}
\right) = k(\varepsilon_{\rm kin})T_{\rm rot}(\beta)
\left(
\begin{array}{c}
{\hat k_x}(\alpha, \beta)	\\
{\hat k_y}(\alpha, \beta)	\\
{\hat k_z}(\alpha, \beta)
\end{array}  
\right). 
\end{eqnarray}
Here, (${\hat k_x}, {\hat k_y}, {\hat k_z}$) is the direction cosine of \mbox{\boldmath $k$} with respect to the basis (\mbox{\boldmath $\bar{e}$}$_x$, \mbox{\boldmath $\bar{e}$}$_y$, \mbox{\boldmath $\bar{e}$}$_z$) fixed to the analyzer's frame, and the effect of the detection type is reflected therein. On the other hand, the effect of the sample orientation is incorporated into $T_{\rm rot}$. The first and second lines of Eq.~(\ref{eq_mapping_sys}) are the forms of $f$ for the angular sector. 

As for the inverse mapping for the angular sector, the problem is reduced to finding the solutions to $\alpha$ and $\beta$ for the following equation:  
\begin{eqnarray}
\label{eq_reorient_sys}
 k(\omega) \left(
	\begin{array}{c}
	{\hat k_x}(\alpha, \beta) \\
	{\hat k_y}(\alpha, \beta) \\
	{\hat k_z}(\alpha, \beta) 
	\end{array}
\right) = T_{\rm rot}^{\text{-}1}(\beta) \left(
\begin{array}{c}
k_x \\
k_y \\
k_z(\omega, k_x, k_y)
\end{array}
\right). 
\end{eqnarray} 

When the analyzer is equipped with a slit-and-deflector, $T_{\rm rot}^{\text{-}1}(\beta)$ does not depend on $\beta$, and the analytic solutions exist owing to the nature of the polar-angular notation; see Eqs.~(\ref{eq_final1}) and (\ref{eq_final2}). 

When the analyzer is equipped with a slit but not a deflector, ${\hat k_{x,y,z}}(\alpha, \beta)$ does not depend on $\beta$, and the existence of the analytic solutions is guaranteed as follows. $T_{\rm rot}^{\text{-}1}(\beta)$ is a rotation matrix. Hence, the entrance condition becomes a linear one-form of $\cos\beta$ and $\sin\beta$. Therefore, the solution to $\beta$ exist in the form $\beta = \tan^{\text{-}1}(A/B)$; see Eqs.~(\ref{eq_inv_theta}) and (\ref{eq_phi_inv_IIa}). Then, from the matching condition, $\alpha$ is solved analytically; see Eqs.~(\ref{eq_inv_alpha}) and (\ref{eq_phi_inv_IIb}). Thus, as long as $T_{\rm rot}^{\text{-}1}(\beta)$ is a rotation matrix, analytic solutions exist. In other words, the existence is guaranteed even when the roto-axis configuration differs from those of types I and II. 

It is thus clarified that, while the analytic forms of $f^{\text{-}1}$ exist for both the slit-type case and slit-less-concept case, the mathematical reasons for the existence differ between the two. The difference originates from whether $T_{\rm rot}^{\text{-}1}$ that describe the rotation of the sample depends on the variable $\beta$ or not, see Eq.~(\ref{eq_reorient_sys}); or in other words, whether the sample is rotated or not during the acquisition of the photoelectron distribution.

\section{Discussion}
\label{sec_discussion}

In early days, analyzers had a hole as the entrance aperture, and band dispersions were tracked by gathering one-dimensional energy distribution curves~\cite{80PRB_Chiang}. Then, analyzers evolved to have a slit~\cite{99Science_Valla}, and more recently, to have a slit-and-deflector so that the concept of the aperture could be removed. The method to manipulate samples also developed considerably. Additional roto-degrees of freedom can be installed by adding a variety of axes in the ultrahigh vacuum~\cite{17RSI_Hoesch,17UltMic_Iwasawa}. 

Each time when the experimental setup is changed, the explicit forms of the mapping function also needs to be updated. This was the first explication of the present article. Second, because the analytic forms of $f^{\text{-}1}$ are guaranteed to exist even for the case when a deflector is adopted (see, section~\ref{sec_systematics}), the mapping program implemented by the algorithm utilizing both $f$ and $f^{\text{-}1}$ can be written for whatever types of the setups. Finally, the datasets recorded at a variety of setups can be handled on equal footings after the systematic nomenclature of the angular variables, as described in section~\ref{sec_systematics}. In Appendix~\ref{appendix}, we summarize the explicit forms of $f$ and $f^{\text{-}1}$ after the nomenclature, and present a demonstrative program that can map the angular distribution onto in-plane momentum space in real time on a standard lap-top computer.

\section*{Supplementary Material}
See Supplementary Material for the demonstrative mapping program that can be loaded on Igor Pro versions 5, 6 and 7. 

\begin{acknowledgments}
The authors acknowledge Peter Baltzer of MB Scientific AB and Karlsson Patrik and Marcus Lundwall of Scienta Omicron for confirming 
that the polar-angular notation is adopted in the deflector-type analyzers commercialized by them; 
%the definition of the angles in the deflector-type analyzers commercialized by them; 
and the anonymous referee for checking that the mapping program can be loaded on Igor Pro version 7. This work was supported by JSPS KAKENHI No.~17K18749. 
\end{acknowledgments}

\appendix

\section{Explicit forms of the mapping functions}
\label{appendix}

\begin{figure} [htb]
\includegraphics[width=8cm]{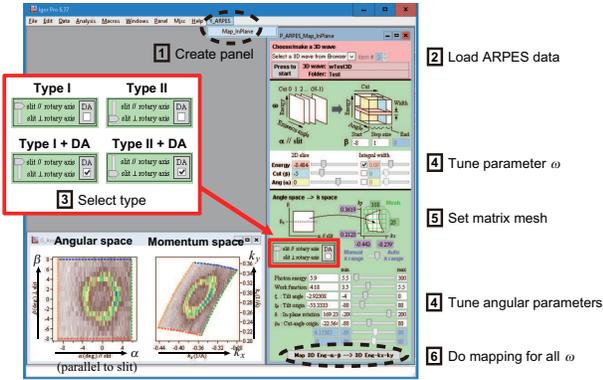}
\caption{\label{fig4} A screen copy of the demonstrative mapping program provided in Supplementary Material~\cite{Suppl}. The panel in the right side of the window is the interface of the program. The left bottom viewgraph in the window displays the ARPES images in angular space (left) and momentum space (right). The boundary of the images in the angular space and momentum space are shown with dotted lines and curves, respectively. }
\end{figure}

We summarize the explicit forms of the mapping and inverse mapping functions considered in the main text. Figure~\ref{fig3} illustrates the setups and the angles. The angles are renamed from those illustrated in Fig.~\ref{fig2} after the nomenclature described in section~\ref{sec_systematics}. In the illustration, we have also introduced new parameters $\beta_0$, $\xi_0$, and $\chi_0$ as the reference to the angles $\beta$, $\xi$, and $\chi$, respectively.

After the renaming of the angles, the two-dimensional angular distribution of photoelectrons $I$ is spanned by the variables ($\alpha$, $\beta$) in all the setups. Such a nomenclature would be useful when writing a program that can map the ARPES datasets recorded under a variety of setups. See the interface panel of a demonstrative program shown in Fig.~\ref{fig4}: Owing to the nomenclature, the angles ($\delta$, $\xi$, $\xi_0$, $\ldots$) always take the role of being tunable parameters irrelevant to which of the four types is selected for the setup.

In principle, the knowledge of $f$ suffice for converting $I(\alpha, \beta)$ onto $k_x$\,-\,$k_y$ plane. Practically, however, knowing the forms of $f^{\text{-}1}$ is useful regarding the computation time for the mapping. The algorithm that utilize both $f$ and $f^{\text{-}1}$ is the following: (1) The boundary of the ARPES data $I(\alpha, \beta)$ is mapped by $f$ onto $k_x$\,-\,$k_y$ plane; see the boundaries indicated by dashed lines/curves on the angular/momentum space shown in Fig.~\ref{fig4}. (2) Thereby, the mapping range in the momentum space is set. (3) A blanc matrix is set in the momentum range with an appropriate mesh size. (4) Use $f^{\text{-}1}$ to refer to the intensity in the angular space from the mesh points $(k_x, k_y)$. A sketch of the algorithm can be seen in the interface panel shown in Fig.~\ref{fig4}. The program also needs a data loading section, and that is located in the upper region of the panel.

Finally, the explicit forms of the in-plane mapping functions $f_i$ for the types $i$ = I, II, I$'$, and II$'$ are, 
\begin{widetext}
\begin{align}
f_{\rm I} &:
\begin{cases} k_x = k \left\{ ({\rm s}\delta{\rm s}\bar{\beta} + {\rm c}\delta {\rm s}\bar{\xi}{\rm c}\bar{\beta}) {\rm c}\alpha - {\rm c}\delta{\rm c}\bar{\xi}{\rm s}\alpha\right\}, \\ k_y = k \left\{ (-{\rm c}\delta{\rm s}\bar{\beta} + {\rm s}\delta {\rm s}\bar{\xi}{\rm c}\bar{\beta}) {\rm c}\alpha - {\rm s}\delta{\rm c}\bar{\xi}{\rm s}\alpha\right\}, \end{cases} \\
f_{\rm II} &:
\begin{cases} k_x = k \left\{ ({\rm s}\delta{\rm s}\bar{\xi} + {\rm c}\delta {\rm s}\bar{\beta}{\rm c}\bar{\xi}) {\rm c}\alpha - ({\rm s}\delta{\rm c}\bar{\xi}{\rm s} - {\rm c}\delta{\rm s}\bar{\beta}{\rm s}\bar{\xi}){\rm s}\alpha\right\}, \\ k_y = k \left\{ (-{\rm c}\delta{\rm s}\bar{\xi} + {\rm s}\delta {\rm s}\bar{\beta}{\rm c}\bar{\xi}) {\rm c}\alpha + ({\rm c}\delta{\rm c}\bar{\xi} + {\rm s}\delta{\rm s}\bar{\beta}{\rm s}\bar{\xi}){\rm s}\alpha\right\}, \end{cases} \\
f_{\rm I'} &:
\begin{cases} k_x = k \{(-\alpha{\rm c}\delta{\rm c}\bar{\xi} + \beta{\rm s}\delta {\rm c}\bar{\chi} - \beta{\rm c}\delta{\rm s}\bar{\xi}{\rm s}\bar{\chi}){\rm sinc}\sqrt{\alpha^2 + \beta^2} + ({\rm s}\delta{\rm s}\bar{\chi} + {\rm c}\delta{\rm s}\bar{\xi}{\rm c}\bar{\chi})\cos\sqrt{\alpha^2 + \beta^2}\}, \\ k_y = k \{(-\alpha{\rm s}\delta{\rm c}\bar{\xi} - \beta{\rm c}\delta {\rm c}\bar{\chi} - \beta{\rm s}\delta{\rm s}\bar{\xi}{\rm s}\bar{\chi}){\rm sinc}\sqrt{\alpha^2 + \beta^2} - ({\rm c}\delta{\rm s}\bar{\chi} - {\rm s}\delta{\rm s}\bar{\xi}{\rm c}\bar{\chi})\cos\sqrt{\alpha^2 + \beta^2}\}, \end{cases} \\
f_{\rm II'} &:
\begin{cases}   k_x = k \{(-\beta{\rm c}\delta{\rm c}\bar{\xi} - \alpha{\rm s}\delta {\rm c}\bar{\chi} + \alpha{\rm c}\delta{\rm s}\bar{\xi}{\rm s}\bar{\chi}){\rm sinc}\sqrt{\alpha^2 + \beta^2} + ({\rm s}\delta{\rm s}\bar{\chi} + {\rm c}\delta{\rm s}\bar{\xi}{\rm c}\bar{\chi})\cos\sqrt{\alpha^2 + \beta^2}\}, \\ k_y = k \{(-\beta{\rm s}\delta{\rm c}\bar{\xi} + \alpha{\rm c}\delta {\rm c}\bar{\chi} + \alpha{\rm s}\delta{\rm s}\bar{\xi}{\rm s}\bar{\chi}){\rm sinc}\sqrt{\alpha^2 + \beta^2} - ({\rm c}\delta{\rm s}\bar{\chi} - {\rm s}\delta{\rm s}\bar{\xi}{\rm c}\bar{\chi})\cos\sqrt{\alpha^2 + \beta^2}\}.  \end{cases} 
\end{align}
%\end{widetext}

Here, $k = 0.513\sqrt{h\nu - \phi_{\rm W} + \omega}$, $\bar{\beta} = \beta - \beta_0$, $\bar{\xi} = \xi - \xi_0$, $\bar{\chi} = \chi - \chi_0$, $\rm{sinc}\,\eta = \sin{\eta} / \eta$, and the types I$'$ and II$'$ refer to those illustrated in Figs.~\ref{fig3}(c) and \ref{fig3}(d), respectively. The angle, energy, and momentum take the units of radian, electron volt, and \AA$^{\text{-}1}$, respectively. The explicit forms of $f_i^{\text{-}1}$ are, 

%\begin{widetext}
\begin{align}
f_{\rm I}^{\text{-}1} &:
\begin{cases} \alpha =& \sin^{\text{-}1}[\{{\rm s}\xi(k^2 - k_x^2 - k_y^2)^{\frac{1}{2}} - {\rm c}\xi({\rm c}\delta k_x + {\rm s}\delta k_y)\}/k], \\ \beta =& \beta_0 + \tan^{\text{-}1}[({\rm s}\delta k_x - {\rm c}\delta k_y) / \{{\rm s}\xi{\rm c}\delta k_x + {\rm s}\xi{\rm s}\delta k_y + {\rm c}\xi(k^2 - k_x^2 - k_y^2)^{\frac{1}{2}}\} ], \end{cases} \\
f_{\rm II}^{\text{-}1} &:
\begin{cases} \alpha =& \sin^{\text{-}1}[\{{\rm s}\xi(k^2 - ({\rm s}\delta k_x - {\rm c}\delta k_y)^2)^{\frac{1}{2}} - {\rm c}\xi({\rm s}\delta k_x - {\rm c}\delta k_y)\}/k], \\ \beta =& \beta_0 + \tan^{\text{-}1}[({\rm c}\delta k_x + {\rm s}\delta k_y) / (k^2 - k_x^2 - k_y^2)^{\frac{1}{2}} ], \end{cases} \\ 
f_{\rm I'}^{\text{-}1} &:
\begin{cases} \alpha =& -\cos^{\text{-}1}[\{t_{31}k_x + t_{32}k_y + t_{33}(k^2 -k_x^2 - k_y^2)^{\frac{1}{2}} \} /k] \\ &\times\{t_{11}k_x + t_{12}k_y + t_{13}(k^2 - k_x^2 - k_y^2)^{\frac{1}{2}} \} / [ k^2 - \{t_{31}k_x + t_{32}k_y + t_{33}(k^2 - k_x^2 - k_y^2)^{\frac{1}{2}}\}^2 ]^{\frac{1}{2}}, \\ \beta =& -\cos^{\text{-}1}[\{t_{31}k_x + t_{32}k_y + t_{33}(k^2 -k_x^2 - k_y^2)^{\frac{1}{2}} \} /k] \\ &\times\{t_{21}k_x + t_{22}k_y + t_{23}(k^2 - k_x^2 - k_y^2)^{\frac{1}{2}} \} / [ k^2 - \{t_{31}k_x + t_{32}k_y + t_{33}(k^2 - k_x^2 - k_y^2)^{\frac{1}{2}}\}^2 ]^{\frac{1}{2}},  \end{cases} \\ 
f_{\rm II'}^{\text{-}1} &:
\begin{cases} \alpha =& \cos^{\text{-}1}[\{t_{31}k_x + t_{32}k_y + t_{33}(k^2 -k_x^2 - k_y^2)^{\frac{1}{2}} \} /k] \\ &\times\{t_{21}k_x + t_{22}k_y + t_{23}(k^2 - k_x^2 - k_y^2)^{\frac{1}{2}} \} / [ k^2 - \{t_{31}k_x + t_{32}k_y + t_{33}(k^2 - k_x^2 - k_y^2)^{\frac{1}{2}}\}^2 ]^{\frac{1}{2}},  \\ \beta =& -\cos^{\text{-}1}[\{t_{31}k_x + t_{32}k_y + t_{33}(k^2 -k_x^2 - k_y^2)^{\frac{1}{2}} \} /k] \\ &\times\{t_{11}k_x + t_{12}k_y + t_{13}(k^2 - k_x^2 - k_y^2)^{\frac{1}{2}} \} / [ k^2 - \{t_{31}k_x + t_{32}k_y + t_{33}(k^2 - k_x^2 - k_y^2)^{\frac{1}{2}}\}^2 ]^{\frac{1}{2}}.  \end{cases}
\end{align}
Here, $t_{ij}$ appearing in the inverse functions of types I$'$ and II$'$ are the elements of $T_{\rm rot}^{\text{-}1}$:
\begin{eqnarray}
T_{\rm rot}^{\text{-}1}  =  \left(
\begin{array}{ccc}
t_{11}	& t_{12}	&	t_{13} \\
t_{21}	& t_{22}	&	t_{23} \\
t_{31}	& t_{32}	&	t_{33} 
\end{array}  
\right) = \left(
\begin{array}{ccc}
\rm{c}\bar{\xi} \rm{c}\delta																				& \rm{c}\bar{\xi} \rm{s}\delta 	&	-\rm{s}\bar{\xi} \\
\rm{s}\bar{\chi} \rm{s}\bar{\xi}\rm{c}\delta	- \rm{c}\bar{\chi} \rm{s}\delta		&	\rm{s}\bar{\chi} \rm{s}\bar{\xi} \rm{s}\delta + \rm{c}\bar{\chi} \rm{c}\delta	&	\rm{s}\bar{\chi} \rm{c}\bar{\xi} \\
\rm{c}\bar{\chi} \rm{s}\bar{\xi}\rm{c}\delta	+ \rm{s}\bar{\chi} \rm{s}\delta	&	\rm{c}\bar{\chi} \rm{s}\bar{\xi} \rm{s}\delta - \rm{s}\bar{\chi} \rm{c}\delta		&	\rm{c}\bar{\chi} \rm{c}\bar{\xi} 
\end{array}  
\right).
\end{eqnarray}

\end{widetext}

The forms could be mistyped in the mapping program. The existence of such an error could be judged by seeing whether the image in the matrix is properly occupying the momentum region set by the boundary; see the boundary of the images shown in Fig.~\ref{fig4}.

\section{Angle notations for the deflector-type analyzer}
\label{appendix_notation}

\begin{figure} [htb]
\includegraphics{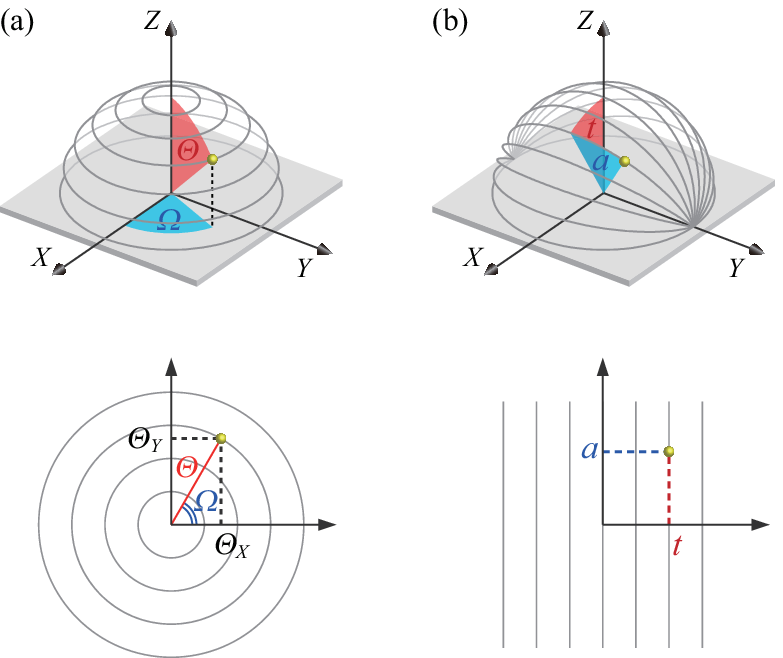}
\caption{\label{fig5} Notations to describe the direction in a hemisphere. (a) Polar-angular notation. (b) Tilt-angular notation. }
\end{figure}

When the deflector-type analyzer is adopted, the direction of the photoelectron is specified by two variables. There is a variety of ways to define the two. Two typical definitions are illustrated in Fig.~\ref{fig5}. In the polar-angular notation [Fig.~\ref{fig5}(a)], the two variables are $\varTheta_X$ and $\varTheta_Y$, and the direction cosine in the Cartesian coordinate $(X, Y, Z)$ is described as $(\sin\varTheta \cos\varOmega, \sin\varTheta \sin\varOmega, \cos\varTheta)$, where  $\varTheta^2 = \varTheta_X^2 + \varTheta_Y^2$ and $\tan\varOmega = \varTheta_Y / \varTheta_X$. In the tilt-angular notation [Fig.~\ref{fig5}(b)], the two are $t$ and $a$, and the direction cosine is $(\cos a \sin t, \sin a, \cos a \cos t)$.

Those who are accustomed to using the slit-type analyzer may be familiar with the tilt-angular notation, because $t$ and $a$ can respectively be regarded as the angle varied step by step and that along the slit direction. Nevertheless, the polar-angular notation is adopted in state-of-the-art deflector-type analyzers~\cite{Communication}. The merit of the polar-angular notation is that a conical photoelectron distribution about the $Z$ axis (constant $\varTheta$) appears as a circular distribution in the $\varTheta_X$-$\varTheta_Y$ plane. In other words, a circular Fermi surface centered at the surface Gamma point appears as a circle in the $\varTheta_X$-$\varTheta_Y$ plane in the normal-emission geometry. This is not the case for the tilt-angular notation: Consider the extreme case $\varTheta$ = 90$^{\circ}$, which appears as a circle in the $\varTheta_X$-$\varTheta_Y$ plane but as two lines at $t = \pm$90$^{\circ}$ in the $t$-$a$ plane. 

If the tilt-angular notation had been adopted in the deflector-type analyzer, there would be one special configuration where the mapping function for the slit type could be used; namely, in the normal-emission geometry where $t$ can be made common to the angle that is varied step by step in the slit-type configuration. However, there is no such chance because the polar-angular notation is adopted in the deflector-type analyzers~\cite{Communication}. Besides, we stress again that, irrelevant to the notation, the forms of the mapping function differ between the slit type and deflector type, in general. Thus, updates in the mapping function are a mandatory when shifting from the slit-type to the deflector-type analyzer. 
 
%\bibliography{ARPES_Geometry_Ref}


\end{document}